\documentclass[12pt, centerh1]{article}

 \usepackage{graphics,arydshln}
 \usepackage{graphicx}

\usepackage{amssymb, natbib}
\usepackage{amsfonts, amsmath, amssymb, marvosym,colonequals}
\usepackage{algorithm,multirow}
\usepackage{bm}
\usepackage{geometry}
\usepackage{lscape}
\usepackage{mathtools}
\usepackage{caption}
\usepackage{theorem}
\usepackage{placeins}
\usepackage{hyperref}
\usepackage{enumerate}
\usepackage{epsfig}

\usepackage{color,soul}

\newcommand{\be}{\begin{equation}}
\newcommand{\ee}{\end{equation}} 
\newcommand{\beq}{\begin{equation*}}
\newcommand{\eeq}{\end{equation*}}
\newcommand{\vecx}{\bm{x}}
\newcommand{\vecX}{\mathbf{X}}

\newcommand{\vecV}{\mathbf{V}}
\newcommand{\vecu}{\bm{u}}
\newcommand{\vecz}{\bm{z}}

\newcommand{\vecvarthet}{\mbox{\boldmath$\vartheta$}}

\newcommand{\load}{\mathbf\Lambda}
\newcommand{\noisev}{\mathbf\Psi}

\newcommand{\vecy}{\mathbf{y}} 
\newcommand{\vecU}{\mathbf{U}}
\newcommand{\vecnu}{\mbox{\boldmath$\nu$}}
\newcommand{\new}{\mbox{\tiny new}}

\newcommand{\vareps}{\mbox{\boldmath$\varepsilon$}}
\newcommand{\vecmu}{\mbox{\boldmath$\mu$}}
\newcommand{\veceta}{\mbox{\boldmath$\eta$}}
\newcommand{\vecalpha}{\mbox{\boldmath$\alpha$}}
\newcommand{\vecSigma}{\mbox{\boldmath$\Sigma$}}

\newcommand{\vecgamma}{\mbox{\boldmath$\gamma$}}

\newcommand{\vectheta}{\mbox{\boldmath$\theta$}}

\newcommand{\vecOmega}{\mbox{\boldmath$\Omega$}}
\newcommand{\vecepsilon}{\mbox{\boldmath$\epsilon$}}
\newcommand{\vecxi}{\mbox{\boldmath$\xi$}}
\newcommand{\veczeta}{\mbox{\boldmath$\zeta$}}

\newcommand{\vecLambda}{\mbox{\boldmath$\Lambda$}}
\newcommand{\vecPsi}{\mbox{\boldmath$\Psi$}}

\newcommand{\matsig}{\mathbf{\Sigma}}

\theorembodyfont{\upshape}

\begin{document}


%
\hyphenation{mcnicholas}
\title{Mixtures of Common Skew-$t$ Factor Analyzers}
\author{Paula M.\ Murray, Paul D.\ McNicholas\thanks{Department of Mathematics \& Statistics, University of Guelph, Guelph, Ontario, N1G 2W1, Canada. E-mail: pmcnicho@uoguelph.ca.} \ and Ryan P.\ Browne}
\date{Department of Mathematics \& Statistics, University of Guelph}
\maketitle

\begin{abstract}

A mixture of common skew-$t$ factor analyzers model is introduced for model-based clustering of high-dimensional data.  By assuming common component factor loadings, this model allows clustering to be performed in the presence of a large number of mixture components or when the number of dimensions is too large to be well-modelled by the mixtures of factor analyzers model or a variant thereof.  Furthermore, assuming that the component densities follow a skew-$t$ distribution allows robust clustering of skewed data. This paper is the first time that skewed common factors have been used, and it marks an important step in robust clustering and classification of high dimensional data.   The alternating expectation-conditional maximization algorithm is employed for parameter estimation. We demonstrate excellent clustering performance when our mixture of common skew-$t$ factor analyzers model is applied to real and simulated data.
\end{abstract}

\section{Introduction}
\label{sec:introduction}

Model-based clustering is an approach to cluster analysis that involves the fitting of a parametric finite mixture model to find groups of similar observations within a data set.  A finite mixture model is of the form 
$f(\vecx\mid\vecvarthet)=\sum_{g=1}^{G}\pi_g f_g(\vecx\mid\vectheta_g),$
such that 
$\pi_g>0$ and $\sum_{g=1}^{G}\pi_g=1$,
where 
$\pi_g$ is the $g$th mixing proportion, $\vectheta_g$ is a vector of parameters, $f_g(\vecx\mid\vectheta_g)$ is the $g$th component density, and $\vecvarthet=(\pi_1,\ldots,\pi_G,\vectheta_1,\ldots,\vectheta_G)$. 
Gaussian model-based clustering has been very popular since Gaussian mixtures were first used for clustering by \cite{wolfe63}. Gaussian mixtures are mathematical tractable and have proven effective in many applications.  However, Gaussian mixtures have limitations, such as difficulty handling skewed data and outliers; accordingly, there has been increased interest in non-Gaussian approaches to mixture modelling of late \citep[e.g.,][]{karlis07,lin09,lin10,montanari10,andrews11a,andrews11c,mcnicholas12,browne12,franczak12,vrbik12,vrbik13,lee12}.

Gaussian mixture models, as well as many non-Gaussian approaches, are not well-suited to modelling data that are high-dimensional due to the prohibitively large number of parameters that must be estimated; for $p$-dimensional data, Gaussian mixtures have $Gp(p+1)/2$ parameters in the component covariance matrices alone. Clustering high-dimensional data continues to prove an important application in the clustering sphere and several approaches to dimension reduction have been developed \citep[e.g.,][]{bouveyron07,mcnicholas08,scrucca10,morris13a,morris13b}.

The mixtures of factor analyzers model \citep[MFA;][]{ghahramani97} is one such approach, and is based on the factor analysis model \citep{spearman04}, which assumes that a $p$-dimensional vector of observed data can be modelled by a $q$-dimensional vector of unobserved factors. The factor analysis model for $p$-dimensional $\vecX$ is: $\vecX=\vecmu+\vecLambda \mathbf{U} + \vecepsilon$, where $\mathbf{U}\sim N(\mathbf{0}, \mathbf{I}_q)$, $\vecepsilon \sim \mathcal{N}(\mathbf{0}, \vecPsi)$,  $\vecLambda$ is a $p\times q$ matrix of factor loadings, and $\vecPsi=$diag$(\psi_1, \psi_2, \ldots, \psi_p)$. Therefore, the density of a MFA model is that of a Gaussian mixture model with component covariance matrices $\vecSigma_g=\vecLambda_g\vecLambda_g'+\vecPsi_g$. Several extensions of the MFA model that include additional restrictions on the component covariance matrices as well as non-Gaussian mixture models have appeared over the past few years \citep[e.g.,][]{mclachlan07,mcnicholas08,andrews11a,murray13}.  

Building on the work of \cite{yoshida04} and \cite{yoshida06}, \cite{baek10} introduced a mixture of common factor analyzers (MCFA) model.  This model assumes that $\vecX$ is modelled as $\vecX=\load\mathbf{U}+\vareps$, where $\load$ is a $p\times q$ matrix, $\vecU\sim \mathcal{N}(\vecxi,\vecOmega)$, $\vareps\sim \mathcal{N}(\mathbf{0},\mathbf{\Psi})$, $\vecxi$ is a $q$-dimensional vector, $\vecOmega$ is a $q\times q$ positive definite symmetric matrix, and $\mathbf{\Psi}$ is a diagonal matrix. The MCFA model places additional restrictions on the component means and covariance matrices compared to the MFA model, thereby further reducing the number of parameters to be estimated.  The density of the MCFA model is
\begin{equation}
f(\vecx)=\sum_{g=1}^{G}\pi_g \phi(\vecx\mid\vecLambda\vecxi_g,\vecLambda\vecOmega_g\vecLambda'+\vecPsi),
\label{eq:mcfa}
\end{equation}
where $\phi(\vecx\mid\vecLambda\vecxi_g,\vecLambda\vecOmega_g\vecLambda'+\vecPsi)$ is the density of the multivariate Gaussian distribution with mean $\vecLambda\vecxi_g$ and covariance matrix $\vecLambda\vecOmega_g\vecLambda'+\vecPsi$. A detailed comparison of the MFA and MCFA methods is given by \cite{baek10}.  One should note that due to the restrictions placed on the component covariance matrices and component location parameters in the MCFA model, the MFA model may be preferable in applications for which it is possible to fit this model --- the same applies to other members of the PGMM family of models \citep{mcnicholas08,mcnicholas10d,mcnicholas10c}, which arises from the imposition of constraints upon the MFA component covariance structure ($\matsig_g=\load_g\load_g'+\noisev_g'$). However, given the need to model data with very high dimensions or a large number of mixture components, the MCFA model may be suitable for clustering data in situations where no member of the PGMM family is sufficiently parsimonious. 

\cite{baek11} subsequently introduced a multivariate-$t$ analogue of the MCFA model referred to as the mixture of common $t$-factor analyzers (MC$t$FA) model. The MC$t$FA has the parsimony advantages of the MCFA model but also accommodates robust clustering. In this work, we develop a mixture of common skew-$t$ factor analyzers (MCS$t$FA) model with an additional restriction on the skewness parameter $\vecalpha_g$, i.e., $\vecalpha_g=\load\veczeta_g$.  This model allows robust clustering of high dimensional data in the presence of skewed data.  

\section{Methodology}
\label{sec:methodology}
The generalized hyperbolic distribution has density function 
\begin{equation}
\begin{split}
h(\vecx\mid\vecvarthet)=& \bigg[\frac{\chi+\delta(\vecx,\vecmu|\vecSigma)}{\psi+\vecalpha'\vecSigma^{-1}\vecalpha}\bigg]^{(\lambda-p/2)/2}
\\&\qquad\qquad\times\frac{[\psi/\chi]^{\lambda/2}K_{\lambda-p/2}\bigg(\sqrt{[\psi+\vecalpha'\vecSigma^{-1}\vecalpha][\chi+\delta(\vecx,\vecmu|\vecSigma)]}\bigg)}{(2\pi)^{p/2}\mid\mathbf\Sigma\mid^{1/2}K_{\lambda}(\sqrt{\chi\psi})\mbox{exp}{(\vecmu-\vecx)'\vecSigma^{-1}\vecalpha}},
\label{eq:ghddensity}
\end{split}
\end{equation}
where $\vecvarthet=(\lambda, \chi, \psi, \vecmu, \vecSigma, \vecalpha)$ is a vector of parameters, and $\delta(\vecx, \vecmu\mid \vecSigma)=(\vecx-\vecmu)'\vecSigma^{-1}(\vecx-\vecmu)$ is the squared Mahalanobis distance between $\vecx$ and $\vecmu$ \citep[cf.][]{mcneil2005}. \cite{browne13} consider a mixture of generalized hyperbolic distributions, i.e., $f(\vecx\mid\vecvarthet)=\sum_{g=1}^{G}\pi_g h(\vecx\mid\vecvarthet_g)$.

The skew-$t$ distribution used herein to develop the MCS$t$FA model arises as a limiting case of the generalized hyperbolic distribution by setting $\lambda=-\nu/2$ and $\chi=\nu$, and letting $\psi\to0$ \citep{barndorff01}. It follows that a $p$-dimensional skew-$t$ random variable $\vecX$ arising from this distribution has density \begin{equation}
\begin{split}
\zeta(\vecx\mid\vecmu,\matsig,\vecalpha,\nu)=&
\bigg[\frac{\nu+\delta(\vecx,\vecmu\mid\vecSigma)}{\vecalpha'\vecSigma^{-1}\vecalpha}\bigg]^{{(-\nu-p)}/{4}}\\
&\quad\qquad\times\frac{\nu^{\nu/2}K_{(-\nu-p)/2}\bigg(\sqrt{[\vecalpha'\vecSigma^{-1}\vecalpha][\nu+\delta(\vecx,\vecmu\mid\vecSigma)]}\bigg)}{(2\pi)^{p/2}\mid\mathbf\Sigma\mid^{1/2}\Gamma({\nu}/{2})2^{\nu/2-1}\exp\{(\vecmu-\vecx)'\vecSigma^{-1}\vecalpha\}},
\label{eq:skewtdensity}
\end{split}
\end{equation} where $\vecmu$ is the location, $\matsig$ is the scale matrix, $\vecalpha$ is the skewness, and $\nu$ is the value for degrees of freedom. Let $\vecX\sim\text{GSt}(\vecmu,\matsig,\vecalpha,\nu)$ represent a skew-$t$ random variable with density in \eqref{eq:skewtdensity}.  By introducing a random variable $Y \sim \Gamma^{-1}(\nu/2,\nu/2)$, where $\Gamma^{-1}(\cdot)$ denotes the inverse Gamma distribution, we can obtain a random variable $\vecX\sim\text{GSt}(\vecmu,\matsig,\vecalpha,\nu)$ through the relationship
$\vecX=\vecmu+Y\vecalpha+\sqrt{Y}\vecV$,
where $\vecV \sim \mathcal{N}(\mathbf{0},\vecSigma)$.  It follows that $\vecX\mid(Y=y)\sim\mathcal{N}(\vecmu+y\vecalpha,y\vecSigma)$ and so, from Bayes' theorem, $Y\mid(\vecX=\vecx)\sim\mbox{GIG}(\vecalpha'\vecSigma^{-1}\vecalpha, \nu+\delta(\vecx,\vecmu\mid\vecSigma),-(\nu+p)/2)$. The generalized inverse Gaussian (GIG) distribution has some attractive features, including tractable expected values. Consider $Y\backsim\text{GIG}(\sqrt{\psi\chi},\sqrt{\chi/\psi},\lambda)$, then the following hold:
\begin{equation}\begin{split}\label{eqn:gig}
&\mathbb{E}[Y] =\sqrt{\frac{\chi}{\psi}} \frac{K_{\lambda+1}(\sqrt{\psi\chi})}{K_\lambda(\sqrt{\psi\chi})},\qquad
\mathbb{E}[1/Y] =\sqrt{\frac{\psi}{\chi}} \frac{K_{\lambda+1}(\sqrt{\psi\chi})}{K_\lambda(\sqrt{\psi\chi})}-\frac{2\lambda}{\chi},\\
&\mathbb{E}[\mbox{log}(Y)] =\mbox{log}\sqrt{\frac{\chi}{\psi}}+\frac{1}{K_\lambda(\sqrt{\psi\chi})}\frac{\delta}{\delta\lambda}K_\lambda(\sqrt{\psi\chi}),
\end{split}\end{equation}
where $\psi, \chi \in \mathbb{R}$ and $K_{\lambda}$ is the modified Bessel function of the third kind with index~$\lambda$.   Extensive details on the GIG distribution and its properties are given by \cite{good53}, \cite{barndorff77}, \cite{blaesild78}, \cite{halgreen79}, and \cite{jorgensen82}.

\cite{murray13} use the representation in \eqref{eq:skewtdensity} to develop a mixture of skew-$t$ factor analyzers model, i.e., a skew-$t$ analogue of the MFA model. Herein, we develop a skew-$t$ analogue of the MCFA model. Recall that the MCFA model assumes that for $\vecX_i$ in component~$g$ we have $\vecX_i=\vecLambda\vecU_{ig}+\vareps$; accordingly, we set $\vecU_{ig}\sim\mathcal{N}(\vecxi_g+Y_i\veczeta_g,Y_i\vecOmega_g)$ and $\vareps\sim\mathcal{N}(\mathbf{0},\mathbf{\Psi})$ to develop a MCS$t$FA model with density
\begin{equation}\label{eqn:ourmodel}
f_{\tiny\text{GSt}}(\vecx\mid\vecvarthet)=\sum_{g=1}^{G}\pi_g\zeta(\vecx\mid\vecLambda\vecxi_g,\load\vecOmega_g\load_g'+\noisev,\load\veczeta_g,\nu_g),
\end{equation}
where $\vecvarthet$ is the vector of all model parameters and the $g$th component is parameterized by location $\vecLambda\vecxi_g$, scale matrix $\load\vecOmega_g\load'+\noisev$, skewness $\load\veczeta_g$, and $\nu_g$ degrees of freedom.

Clearly, our MCS$t$FA model \eqref{eqn:ourmodel} has its place as an asymmetric model-based approach to clustering, classification, and discriminant analysis. However, there are already several other approaches in that category \citep[e.g.,][]{karlis09,lin09,montanari10,vrbik12,franczak12,ho12,lin13,browne13}, and it is only through an appreciation of the extent of its parsimony that one can truly grasp the role of our MCS$t$FA model within the wider pallet of mixture modelling approaches. Consider the parsimonious mixture of skew-$t$ factor analyzers (MS$t$FA) family introduced by \cite{murray13}. The MS$t$FA family comprises eight models (Table~\ref{tab:models}) that arise from the imposition of constraints on the scale matrix of the most general member, which has a similar density to the model in \eqref{eqn:ourmodel} except that the location is $\vecmu_g$ rather than $\vecLambda\vecxi_g$ and the scale matrix is $\vecLambda_g\vecLambda_g'+\vecPsi_g$ rather than $\vecLambda\vecOmega_g\vecLambda+\vecPsi$.
\begin{table}[h]
\caption{The number of free parameters and nomenclature for each member of the PMSTFA family, and the number of free parameters for our MCS$t$FA model.}
\begin{tabular*}{\textwidth}{@{\extracolsep{\fill}}llrrr}
\hline
\multicolumn{2}{c}{Family/Model}  & $\matsig_g$ & Free Model Parameters\\
 \hline  
\multirow{8}{*}{PMSTFA}&CCC&$\vecLambda\vecLambda'+\psi\mathbf{I}_p$&$pq-q(q-1)/2+2Gp+2G$\\
&CCU&$\vecLambda\vecLambda'+\vecPsi$&$pq-q(q-1)/2+2Gp+2G+p-1$\\
&CUC&$\vecLambda\vecLambda'+\psi_g\mathbf{I}_p$&$pq-q(q-1)/2+2Gp+3G-1$\\
&CUU&$\vecLambda\vecLambda'+\vecPsi_g$&$pq-q(q-1)/2+3Gp+2G-1$\\
&UCC&$\vecLambda_g\vecLambda_g'+\psi\mathbf{I}_p$&$G[pq-q(q-1)/2]+2Gp+2G$\\
&UCU&$\vecLambda_g\vecLambda_g'+\vecPsi$&$G[pq-q(q-1)/2]+2Gp+2G+p-1$\\
&UUC&$\vecLambda_g\vecLambda_g'+\psi_g\mathbf{I}_p$&$G[pq-q(q-1)/2]+2Gp+3G-1$\\
&UUU&$\vecLambda_g\vecLambda_g'+\vecPsi_g$&$G[pq-q(q-1)/2]+3Gp+2G-1$\\
\hdashline
MCS$t$FA&&$\vecLambda\vecOmega_g\vecLambda+\vecPsi$&$Gq(q+1)/2+q(p+2G-q)+2G+p-1$\\
\hline
\end{tabular*}
\label{tab:models}
\end{table}

Because of the \textit{prima facie} similarity between the members of the MS$t$FA family and our MCS$t$FA model, one may be apt to underestimate the difference in their respective levels of parsimony. Heretofore, the MS$t$FA family contained the most parsimonious non-elliptical model-based approaches to clustering, classification, and discriminant analysis within the literature. Upon careful consideration of the rightmost column of Table~\ref{tab:models}, one can get a sense of a difference in parsimony. For members of the MM$t$FA family, the term involving $p$ is of the form $(aq+bG)p$, for $a\in\{1,G\}$ and $b\in\{2,3\}$; however, for our MCS$t$FA model, the equivalent term is $(q+1)p$, which is necessarily smaller. 

\begin{figure}[h!]
\vspace{-0.2in}
\centering
\includegraphics[height=3.1in]{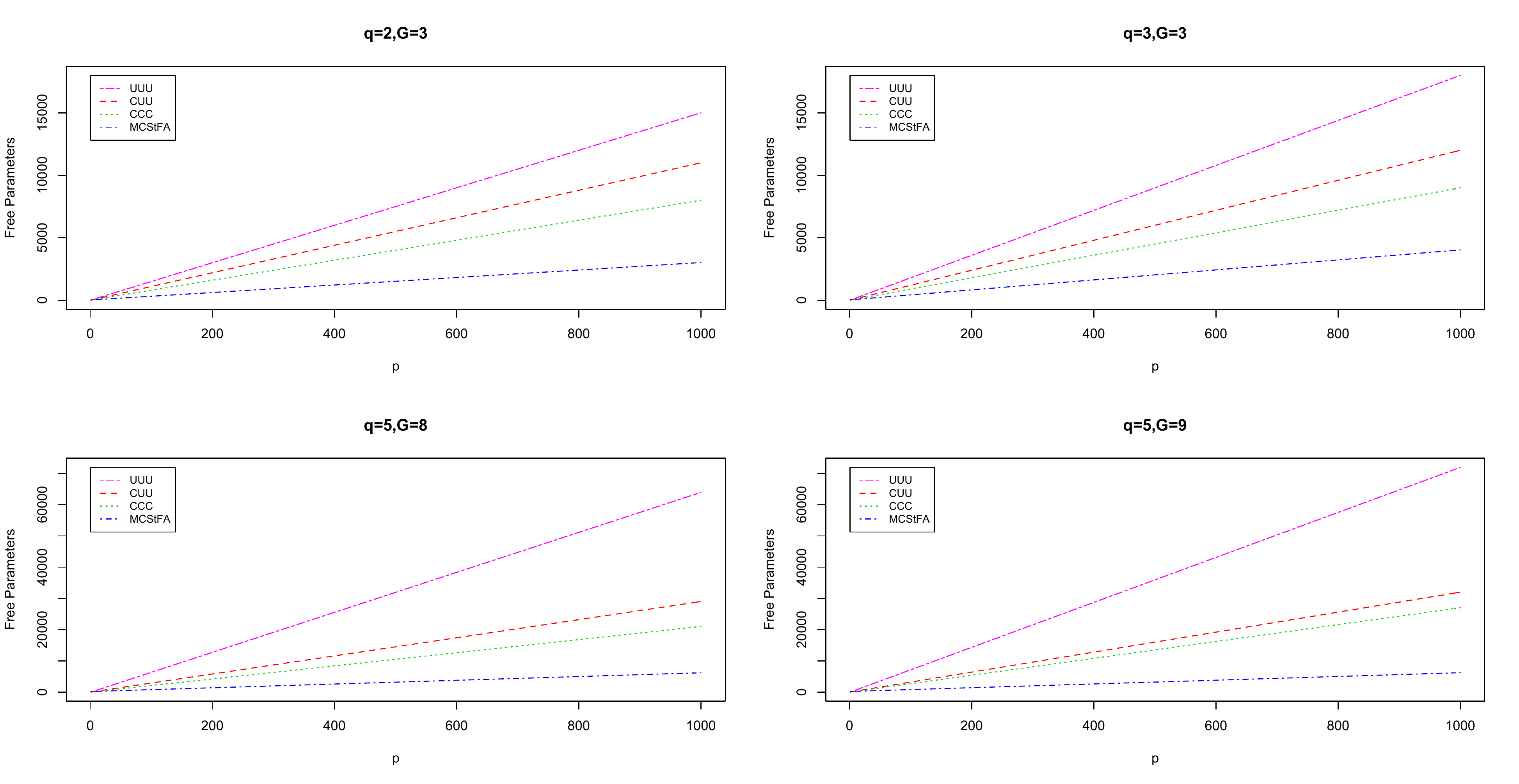}
\vspace{-0.2in}
  \caption{Scatter plots illustrating how the number of free parameters grows with data dimensionality $p$, using four values of $(q,G)$, for three members of the MS$t$FA family (UUU, CUU, CCC) and our MCS$t$FA model.}
  \label{fig:paras}
\end{figure}

To bring this point into sharper relief, consider some plots (Figure~\ref{fig:paras}) of the number of free parameters for three members of the MS$t$FA family (UUU, CUU, and CCC) and our MCS$t$FA model. Note that the UUU model is the least parsimonious MS$t$FA model and the CCC model is the most parsimonious --- although, the latter would rarely be useful in practice because it imposes common scale matrices across components. The CUU model has common factor loading matrices and is more parsimonious than the UUU model but more flexible than the CCC model. In the first row of Figure~\ref{fig:paras}, we fix $G=3$ and consider how the number of free parameters grows with $p$ for $q=2$ and $q=3$, respectively. The second row is similar except that now we fix $q=5$ and consider how the number of free parameters grows with $p$ for $G=8$ and $G=9$, respectively. From the plots in Figure~\ref{fig:paras}, it is very clear that our MCS$t$FA model is much more parsimonious than any MS$t$FA model, and that the extent of the difference in parsimony between our MCS$t$FA model and the MS$t$FA models grows with $p$. The fact the the number of free parameters grows quite gently with $p$ --- especially when one compares to the MS$t$FA models --- is an attractive feature of our MCS$t$FA model.

\section{Parameter Estimation}
\subsection{AECM Algorithm}
The alternating expectation-conditional maximization (AECM) algorithm \citep{meng97} was employed for parameter estimation in an analogous fashion to the algorithm described by \cite{baek11}.  The E-step requires the computation of the expectations in \eqref{eqn:gig} as well as the expected value of the component membership labels, 
$$\mathbb{E}[Z_{ig}|\vecx_i] = \frac{\pi_g f(\vecx_i|\vectheta_g)}{\sum^{G}_{h=1}\pi_hf(\vecx_i|\vectheta_h)}=:\hat{z}_{ig},$$ 
where $Z_{ig}$ indicates component membership and is defined so that $z_{ig}=1$ if $\vecx_i$ is in component~$g$ and $z_{ig}=0$ otherwise. We also need the following conditional expectations:
\begin{equation*}\begin{split}
&\mathbb{E}[\vecU_{ig}-\vecxi_g\mid\vecx_i,y_i,Z_{ig}=1]=\vecgamma_g'(\vecx_i-\vecLambda\vecxi_g),\\
&\mathbb{E}[(\vecU_{ig}-\vecxi_g)(\vecU_{ig}-\vecxi_g)'\mid\vecx_i,y_i,Z_{ig}=1]=\vecgamma_g'(\vecx_i-\vecLambda\vecxi_g)(\vecx_i-\vecLambda\vecxi_g)'\vecgamma_g\\&\qquad\qquad\qquad\qquad\qquad\qquad\qquad\qquad\qquad\qquad\qquad\qquad\qquad\qquad+y_i(\mathbf{I}_q-\vecgamma_g'\vecLambda)\vecOmega_g,
\end{split}\end{equation*}
where $\vecgamma_g=\left(\load\vecOmega_g\load'+\vecPsi\right)^{-1}\load\vecOmega_g$.

Within the framework of the AECM algorithm,  the complete-data comprise the observed data $\vecx_i$, the group membership labels $z_{ig}$, the latent $y_{ig}$, and the latent factors $\vecu_{ig}$, for $i=1,\ldots,n$ and $g=1,\ldots,G$. Therefore, the complete-data log-likelihood is
\begin{equation*}\begin{split}
l_c(\vecvarthet\mid\vecx,\vecy,&\vecz,\vecu)=\sum^{n}_{i=1}\sum^{G}_{g=1}z_{ig}\big[\mbox{log} \pi_g+\mbox{log} \phi(\vecx_i\mid\vecLambda\vecu_{ig},y_i\mathbf{\Psi})\\&\qquad\qquad\qquad\qquad+\mbox{log} \phi(\vecu_i\mid\vecxi_g+y_i\veczeta_g,y_i\vecOmega_g)+\mbox{log} h(y_i\mid\nu_g/2,\nu_g/2)\big].\end{split}\end{equation*}

On the first conditional-maximization (CM) step, the missing data are taken to be the group membership labels $z_{ig}$ and the latent $y_{ig}$. At this stage, we compute the following parameter updates:
$$\hat{\pi}_g= \frac{n_g}{n}, \quad \hat{\vecxi}_g=\frac{ 1}{m_g}\sum^{n}_{i=1}(\vecgamma_g'\vecLambda)^{-1}\vecgamma_g'\vecx_i( \overline{a}_g b_{ig}-1),
\quad\text{and}\quad\hat{\veczeta}_g=\frac{1}{m_g}\sum_{i=1}^{n}\vecgamma_g'\vecx_i(\overline{b}_g-b_{ig}),$$
and we solve the equation 
$$\mbox{log}\bigg(\frac{\hat{\nu}^{\new}_g}{2}\bigg)+1-\varphi\bigg(\frac{\hat{\nu}^{\new}_g}{2}\bigg)-\sum^{n}_{i=1}\bigg(z_{ig}c_{ig}+b_{ig}\bigg)=0$$ for $\nu_g^{\new}$, numerically, to obtain the update for $\nu_g$. Note that:
\begin{itemize}
\item the expected values $a_{ig}=\mathbb{E}[Y_i|\vecx_i,Z_{ig}=1] $, $b_{ig} =\mathbb{E}[1/Y_i|\vecx_i,Z_{ig}=1]$, and $c_{ig} =\mathbb{E}[\mbox{log}(Y_i)|\vecx_i,Z_{ig}=1]$ are computed using the results in \eqref{eqn:gig}, and
\item $n_g=\sum^{n}_{i=1}z_{ig}$, $\overline{a}_g= ({1}/{n_g})\sum^{n}_{i=1}z_{ig}a_{ig}$, $ \overline{b}_g= ({1}/{n_g})\sum_{i=1}^{n}z_{ig}b_{ig}$, and $m_g = \sum_{i=1}^{n} \overline{a}_gb_{ig}-n_g$.
\end{itemize}

On the second CM-step, the missing data consist of the group membership labels $z_{ig}$, the latent $y_i$, and the latent factors $\vecu_{ig}$. We estimate the common factor loading matrix $\vecLambda$, the factor covariance parameters $\vecOmega_1,\ldots,\vecOmega_G$, and the diagonal matrix $\mathbf{\Psi}$. The updates for $\vecLambda$, $\mathbf{\Psi}$, and $\mathbf{\Omega}_g$ are given by 
\begin{equation*}\begin{split}
\hat{\vecLambda} &= \bigg( \sum^{G}_{g=1}\sum^{n}_{i=1}z_{ig}a_{ig}x_i\veceta_{ig}'  \bigg)\left\{\sum^{G}_{g=1}\left(\sum^{n}_{i=1}z_{ig}a_{ig}\veceta_{ig}\veceta_{ig}'+n_g\left(\mathbf{I}_q-\vecgamma_g'\vecLambda\right)\vecOmega_g\right)\right\}^{-1},\\ 
\hat{\mathbf{\Psi}} &= \frac{1}{\sum^{G}_{g=1}n_g}\sum^{G}_{g=1}n_g\left\{\left(\hat{\vecLambda}\vecgamma_g'-\mathbf{I}_p\right)\mathbf{S}_g\left(\hat{\vecLambda}\vecgamma_g'-\mathbf{I}_p\right)'+\hat{\vecLambda}\vecOmega_g\left(\mathbf{I}_q-\hat{\vecLambda}'\vecgamma_g\right)\hat{\vecLambda}'\right\},\\ 
\hat{\mathbf{\Omega}}_g &=\vecgamma_g'\mathbf{S}_g\vecgamma_g+\vecOmega_g\left(\mathbf{I}_q-\hat{\vecLambda}'\vecgamma_g\right) ,
\end{split}\end{equation*} 
where $${\mathbf{S}}_g=\frac{1}{n_g}\sum_{i=1}^{n}\hat{z}_{ig}b_{ig}(\vecx_i-\hat{\vecLambda}\hat{\vecxi}_g)(\vecx_i-\hat{\vecLambda}\hat{\vecxi}_g)'-\hat{\vecLambda}\hat{\veczeta}_g(\bar{\vecx}_g-\hat{\vecLambda}\hat{\vecxi}_g)'-(\bar{\vecx}_g-\hat{\vecLambda}\hat{\vecxi}_g)(\hat{\vecLambda}\hat{\veczeta}_g)'+\overline{a}_g\hat{\vecLambda}\hat{\veczeta}_g(\hat{\vecLambda}\hat{\veczeta}_g)'$$ and
$\veceta_{ig} = \vecgamma_g'\left(\vecx_i-\vecLambda\hat{\vecxi}_g\right)+\hat{\vecxi}_g$.

Convergence of our AECM algorithms is determined using a criterion based on the Aitken acceleration \citep{aitken26}. Specifically, the Aitken acceleration can used to estimate the asymptotic maximum of the log-likelihood at each iteration of an EM algorithm and thence to determine convergence. The Aitken acceleration at iteration $t$ is 
$$a^{(t)}=\frac{l^{(t+1)}-l^{(t)}}{l^{(t)}-l^{(t-1)}},$$
where $l^{(t)}$ is the log-likelihood at iteration $t$. An asymptotic estimate of the log-likelihood at iteration $t+1$ is
$$l_{\infty}^{(t+1)}=l^{(t)}+\frac{1}{1-a^{(t)}}(l^{(t+1)}-l^{(t)}),$$
and the algorithm can be considered to have converged when
$l_{\infty}^{(t)}-l^{(t)}<\epsilon$ \citep{bohning94,lindsay95}.

\subsection{Model Selection and Performance Assessment}

The Bayesian information criterion \citep[BIC;][]{schwarz78} is used to select number of mixture components and the number of latent factors. The BIC is defined as $\text{BIC}=2l(\vecx,\hat{\vecvarthet})- \rho$ log $n$, where $l(\vecx,\hat{\vecvarthet})$ is the maximized log-likelihood, $\hat{\vecvarthet}$ is the maximum likelihood estimate of the model parameters $\vecvarthet$, $\rho$ is the number of free parameters in the model, and $n$ is the number of observations. Support for the use of the BIC in mixture model selection is given by \cite{campbell97} and \cite{dasgupta98}, while \cite{lopes04} provide support for its use in selecting the number of latent factors in a factor analysis model.

We carry out cluster analysis on data sets with known classes to assess the performance of the MCS$t$FA model in capturing the underlying classes. Therefore, we can use the adjusted Rand index \citep[ARI;][]{hubert85} to measure the class agreement between the true and estimated group memberships.  An ARI value of 1 indicates perfect class agreement, a value of 0 indicates results that would be expected under random classification, and a negative value indicates classification that is worse than would be expected by chance.

\section{Illustrations}
\label{sec:applications}

\subsection{Initialization}

For all analyses performed herein, the MCS$t$FA model is applied from agglomerative hierarchical clustering starting values obtained using the $\tt{hclust}$ function in the $\sf{R}$ software \citep{R}.  The initial degrees of freedom are set as $\nu_g^{(0)}$=50 and the skewness parameters are initialized as $p$-dimensional vectors with all entries equal to $1$.  The initializations of all other model parameters are analogous to those used by \cite{baek10} in the Gaussian case.
For comparison, we also fit the MC$t$FA model using the {\tt mixfa} package \citep{rathnayake13} for {\sf R} with the same starting values. The reason we use the MC$t$FA for comparison is to illustrate that our model, with the added ability to model skewness, is worthwhile above and beyond the MC$t$FA model. 

\subsection{Simulation Study}

To illustrate the clustering ability of the MCS$t$FA model, a $p=15$ dimensional data set is simulated from a $G=4$ component MCS$t$FA model with $q=2$ latent factors.  The values of the factor loading matrix $\vecLambda$ are simulated from the standard normal distribution. The degrees of freedom are set at $\vecnu=(5,2,40,40)$ and the skewness parameters of the latent factors $\vecU_{ig}$ are set as $\veczeta_1=(10, 10)'$,  $\veczeta_2=(0, 0)'$, $\veczeta_3=(0, 0)'$, and $\veczeta_4=(50, 45)'$, respectively. Data are simulated for $n=200$ observations with $\pi_g=1/4$ for all $g$. The MCS$t$FA model is fitted to these data for $q=1,\ldots,10$ latent factors and $G=4$ components. The model with $q=2$ latent factors obtains the highest BIC value ($-12970.96$) and gives perfect clustering results ($\text{ARI}=1.00$).  A comparison of the true and predicted classifications is given in Table~\ref{tab:simulation}.
\begin{table}[h]
\caption{A cross-tabulation of true (A,B,C,D) and predicted (1,2,3,4) group memberships for the selected MCS$t$FA and MC$t$FA models, respectively, for the simulated data.}
\centering
\begin{tabular*}{1.0\textwidth}{@{\extracolsep{\fill}}lccccccccc}
\hline
 &\multicolumn{4}{c}{MCS$t$FA}&&\multicolumn{4}{c}{MC$t$FA}\\
\cline{2-5}	\cline{7-10}	
& 1 & 2 & 3 & 4 && 1 & 2 & 3 & 4\\ 
  \hline  
A & 50 &  &  & &&  50&  & & \\
B & & 50 & & &&  & 50& &\\
C & & & 50 &  && & 50 & &\\
D & & & & 50 && & & 6 &44\\
\hline
\end{tabular*}
\label{tab:simulation}
\end{table}

\begin{figure}[h!]
\vspace{-0.2in}
\centering
\includegraphics[width=4in,height=3.1in]{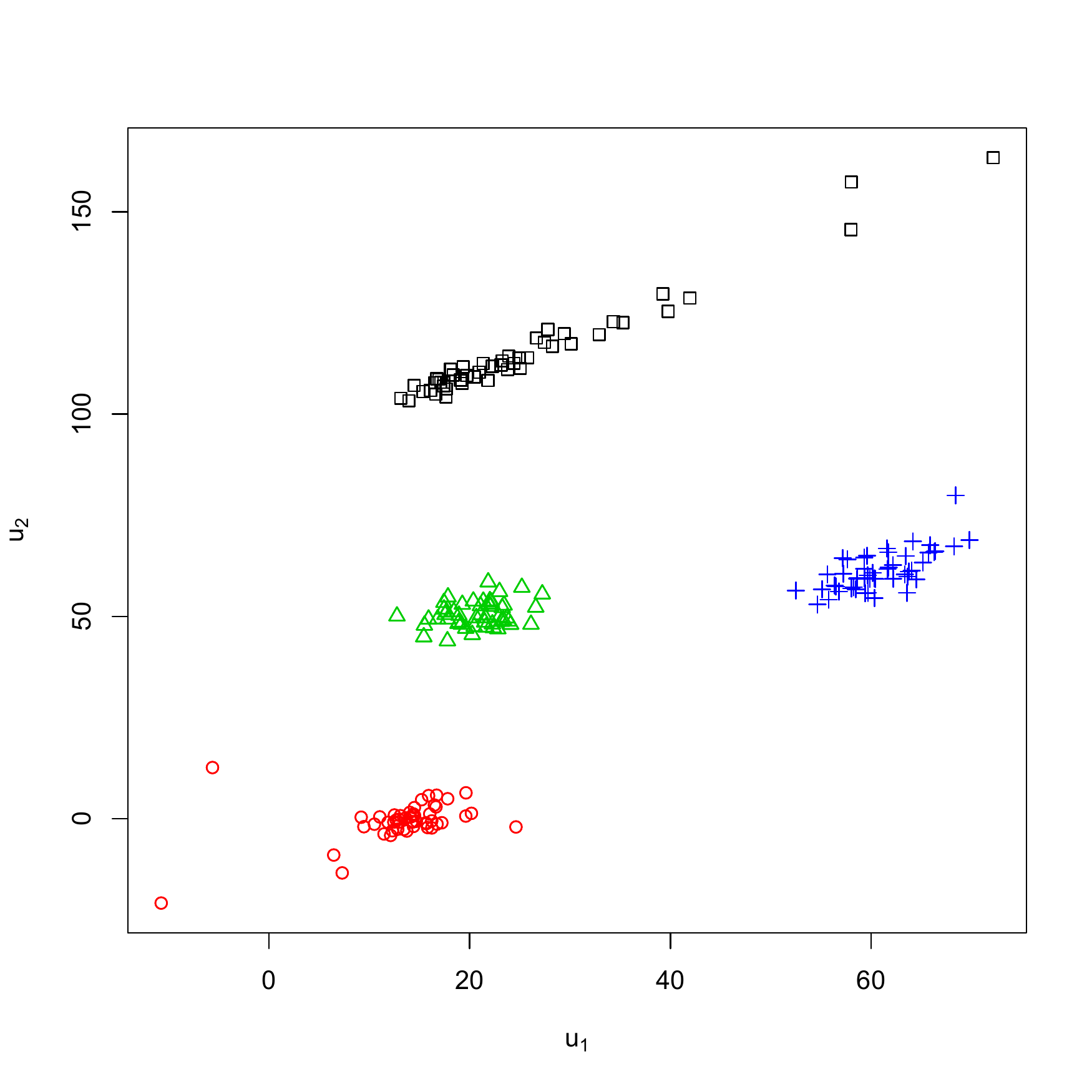}
\vspace{-0.2in}
  \caption{Scatter plot illustrating the clustering results in the latent common factor space for the MCS$t$FA model with $q$=2 latent factors for the simulated data.}
  \label{fig:simplot}
\end{figure}

Figure \ref{fig:simplot} shows a plot of the latent factors where colour corresponds to predicted classification and shape corresponds to true group membership.  Note that from their parameters, we can see that these components could be described by four different statistical distributions: multivariate Gaussian, multivariate-$t$, multivariate skew-$t$, and multivariate skew-normal.  These clustering results demonstrate the flexibility of the MCS$t$FA model in fitting data effectively arising from different underlying distributions that are special cases of the skew-$t$.  

\begin{figure}[h!]
\vspace{-0.2in}
\centering
\includegraphics[width=4in,height=3.1in]{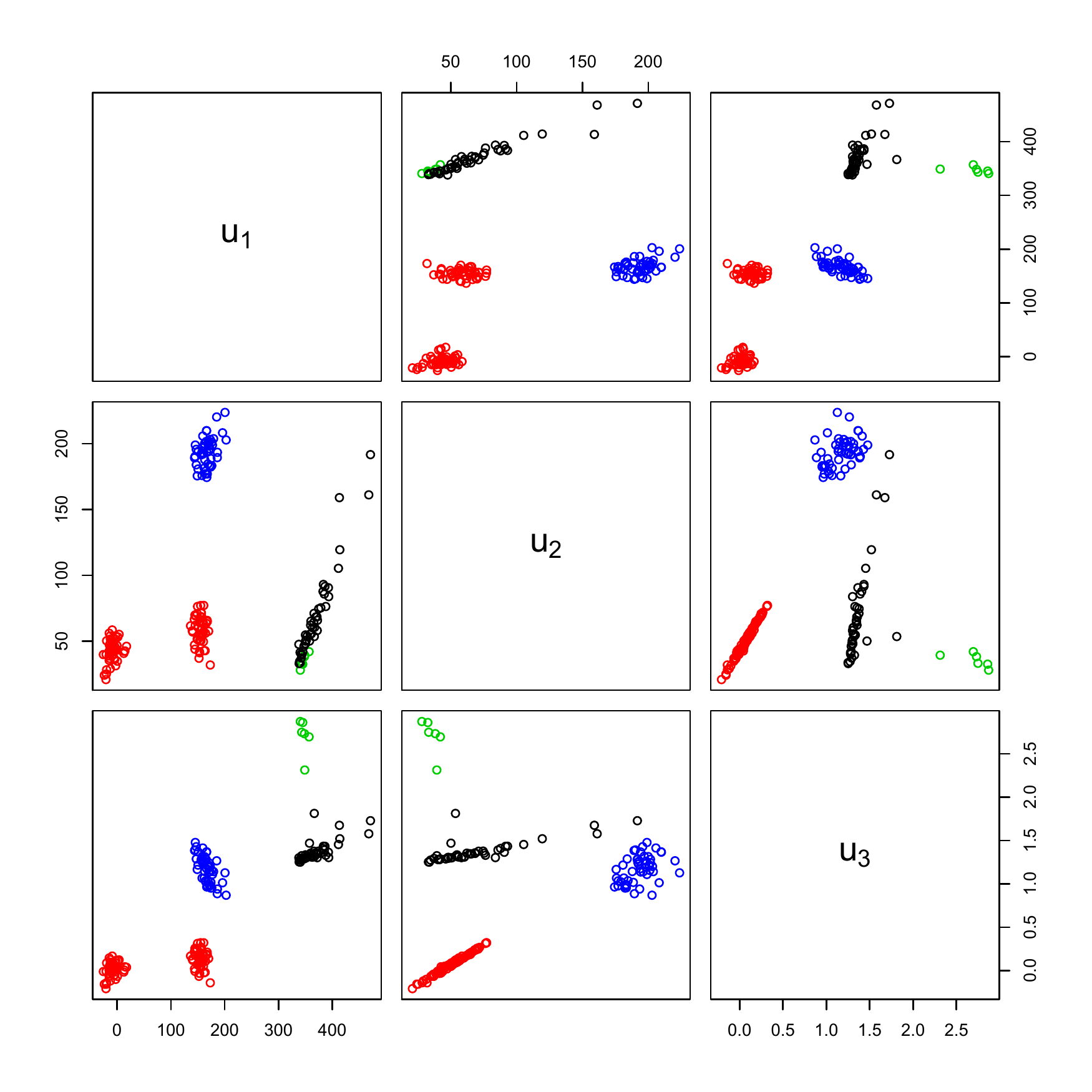}
\vspace{-0.2in}
  \caption{Scatter plot illustrating the clustering results in the latent common factor space for the MC$t$FA model with $q$=3 latent factors for the simulated data.}
  \label{fig:simplot2}
\end{figure}

For comparison, the MC$t$FA model is also fit to these simulated data using the same starting values. 
In this case, the best model in terms of BIC ($-12179.31$) is the model with $q=3$ latent factors and $G=4$ components (Figure~\ref{fig:simplot2}). We can see (Figure~\ref{fig:simplot2}) that the clustering results are essentially determined by the second and third latent common factors. The classifications (Table~\ref{tab:simulation}) resulting from this model give an ARI value of $0.68$. 

\subsection{Breast and Colon Cancer Data}

We apply the MCS$t$FA models to gene expression data from 32 matched breast tumor tissue pairs and 20 matched colon tissue pairs \citep{chowdary06}. We use the reduced version of this data set containing 182 gene expression values for each sample \citep{souto08}.  
We fit the MCS$t$FA model to the data for $G=2$ components and $q=1,\ldots,9$ latent factors.  The model with $q=3$ latent factors obtains the highest BIC value ($-35930.61$) with an associated ARI value of 0.92.  A cross-tabulation of the true and predicted classifications is given in Table \ref{tab:chowdary}.
\begin{table}[h]
\caption{A cross-tabulation of true and predicted group memberships for the selected MCS$t$FA model for the breast and colon cancer data.}
\centering
\begin{tabular*}{\textwidth}{@{\extracolsep{\fill}}lcc}
\hline
 &\multicolumn{2}{c}{Predicted}\\
\cline{2-3}	
& 1 & 2 \\ 
  \hline  
Breast\ Tumour & 61 & 1 \\
Colon\ Tumour & 1 & 41 \\
\hline
\end{tabular*}
\label{tab:chowdary}
\end{table}

We choose these data for analysis herein because they were previously analyzed by \cite{baek11}.  In their analysis, the MC$t$FA model is fit to the data for $G=2$ components and $q=1,\ldots,9$ latent factors from 50 starting values for the parameters. Parameter estimation is carried out in a very similar fashion to the method used for our analysis.  Note that \cite{baek11} obtain a best ARI of 0.89 with a MC$t$FA model with $q=6$ latent factors and $G=2$ components.  While the MC$t$FA model performs well on these data, the additional flexibility provided by the skewness parameter in the MCS$t$FA leads to superior classification results.
\begin{figure}[h!]
\vspace{-0.34in}
\centering
\includegraphics[width=2.8in]{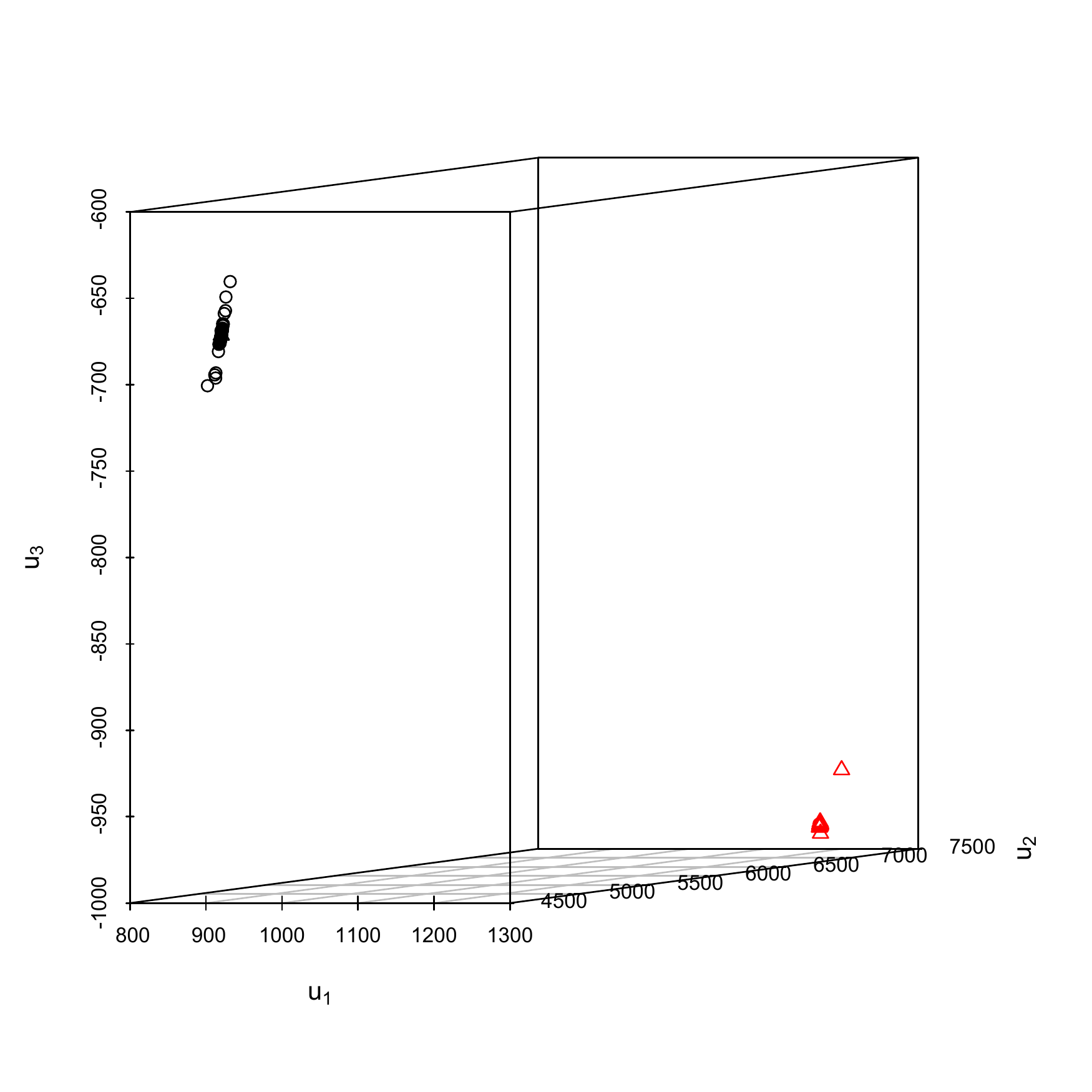}
\vspace{-0.25in}
  \caption{Scatter plot illustrating the clustering results for the best MCS$t$FA model for the breast and colon cancer data.}
  \label{fig:ChowdaryPlot}
\end{figure}

\subsection{Leukaemia Data}

\cite{armstrong02} introduced gene expression data corresponding to three subclasses of leukaemia: lymphoblastic leukaemia with MLL translocations (MLL), conventional acute lymphoblastic leukaemia (ALL), and acute myelogenous leukaemia (AML).  The data contain gene expression measurements from 20, 24, and 28 samples from these three classes, respectively.  \cite{souto08} subsequently introduced a reduced version of this data set containing 2194 gene expression values per sample.  Prior to our analysis, the data were log-scaled and genes were eliminated if the maximum log-scaled expression value across all samples was less than 3.5 times the minimum value.  After gene-filtering, 552 genes remained and clustering of the samples was carried out on the basis of these genes.  The MCS$t$FA model was fit to the data for $q=1,\ldots,10$ latent factors with $G=3$ components.  The best model in terms of BIC ($-61136.22$) was a $q=1$ factor model with an ARI value of 0.74.  Table~\ref{tab:Armstrong1} gives a comparison of the true and estimated group memberships. 
\begin{table}[h]
\caption{A cross-tabulation of true and predicted group memberships for the selected MCS$t$FA and MC$t$FA model, respectively, for the leukaemia data.}
\centering
\begin{tabular*}{\textwidth}{@{\extracolsep{\fill}}lccccccc}
\hline
 &\multicolumn{3}{c}{MCS$t$FA}& &\multicolumn{3}{c}{MC$t$FA}\\
\cline{2-4}	\cline{6-8}	
& 1 & 2 & 3&& 1 & 2 & 3\\ 
  \hline  
MLL & 20 &  & && 20 & &\\
ALL  & 4 & 19 & 1 && 6 & 17 & 1 \\
AML & & 2 & 26 && & 1 & 27 \\
\hline
\end{tabular*}
\label{tab:Armstrong1}
\end{table}

\begin{figure}[h!]
\centering
\vspace{-0.2in}
\includegraphics[width=2.85in,height=2.65in]{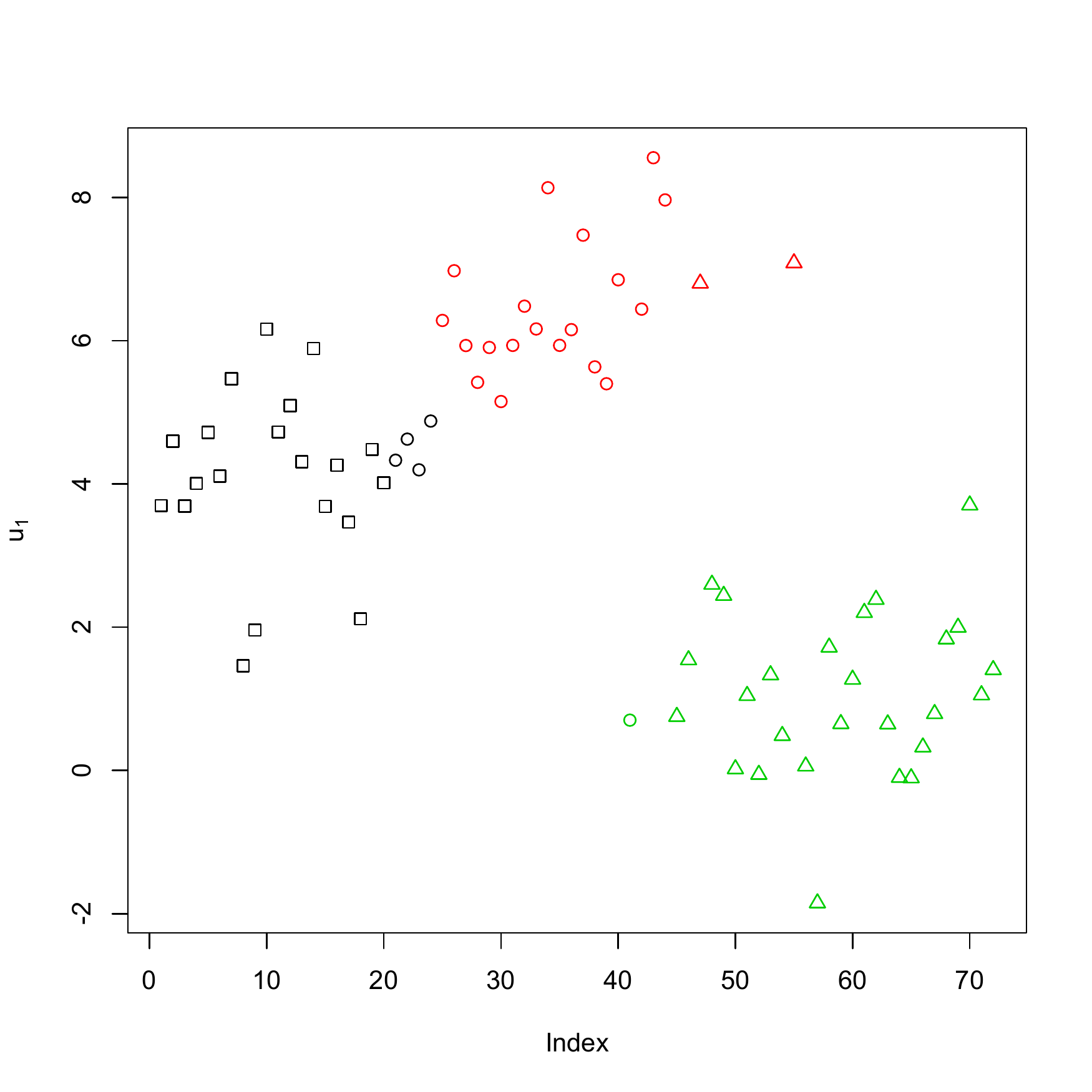}
\includegraphics[width=2.85in,height=2.65in]{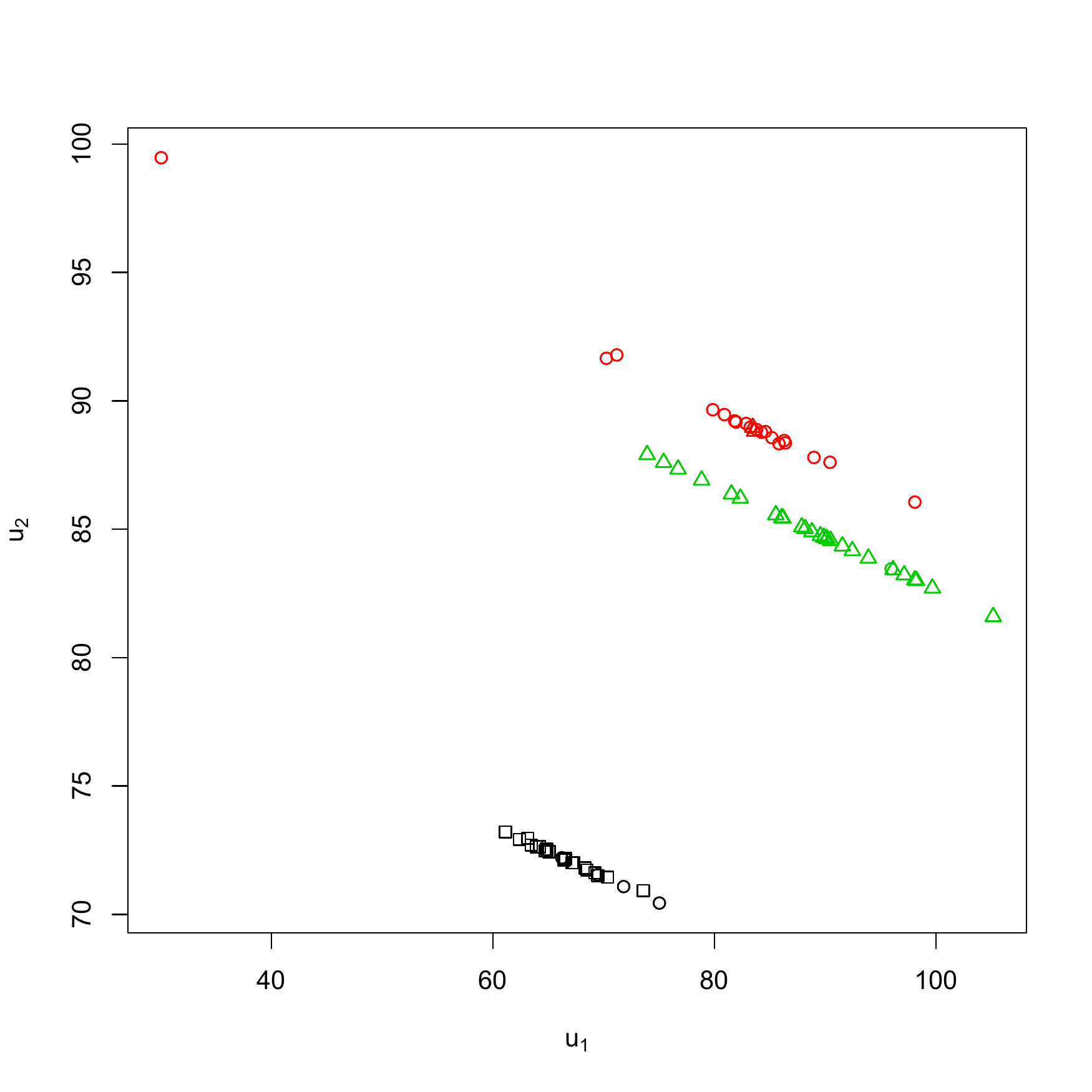}
\vspace{-0.12in}
  \caption{Scatter plots illustrating the clustering results for the MCS$t$FA models with $q=1$ and $q=2$ latent factors, respectively, for the leukaemia data; shapes represent true labels and colours denote predicted classifications.}
  \label{fig:armstrong2}
\end{figure}

For illustration purposes, a plot of our $q=1$ common factor solution is given in Figure~\ref{fig:armstrong2}; the next most parsimonious model has $q=2$ common factors, gives exactly the same clustering results, and is also shown in Figure~\ref{fig:armstrong2}.

The MCtFA model was also fit to the data with the best model in terms of BIC ($-158679.7$) was a $q=5$ factor model ($\text{ARI}=0.72$).  Table \ref{tab:Armstrong1} gives a cross-tabulation of the true group memberships and the cluster memberships as predicted by the MC$t$FA model. Note that the estimated skewness parameters of the latent factors in our MCS$t$FA model were near zero, i.e., $\veczeta=\left(1.58\times{10}^{-12},3.34\times{10}^{-12},2.03\times{10}^{-12}\right)$; therefore, we would expect similar clustering results from the MCS$t$FA and MC$t$FA models. 

Note that for these data, the estimated degrees of freedom obtained by our MCS$t$FA model were very low, i.e., $\vecnu=(1.3,1.4,2.0)$. Some may prefer that the restriction $\nu>6$ is imposed, so that the first three moments will exist; while we do not think that this is necessary in practice, we can confirm that identical clustering results are achieved if this restriction is imposed here.

\section{Summary}\label{sec:conc}

We have introduced a MCS$t$FA model based on a form of the skew-$t$ distribution that arises as a special case of the generalized hyperbolic distribution. Through this representation, an elegant and mathematically tractable skew-$t$ common factor analyzers model arises. Borrowing attractive features of the GIG distribution, the form of the skew-$t$ distribution we use lends itself to elegant parameter estimation via an AECM algorithm. Previous work has focused on mixtures of skew-normal factor analyzers \citep{montanari10,lin13} and mixtures of skew-$t$ factor analyzers \citep[i.e., the MM$t$FA family;][]{murray13}; however, this paper is the first instance of a common factors mixture model that accounts for skewness. Because we have shown examples where our MCS$t$FA model outperforms its symmetric analogue (MC$t$FA), it is clear that parameterizing skewness can lead to an advantage in applications of common factors. 

Our approach is analogous to the development of a mixture of skew-$t$ factor analyzers model by \cite{murray13} and future work will consider extensive comparison of the approaches. Although \cite{murray13} consider a family of eight parsimonious models, all with a number of covariance parameters that is linear in data dimensionality, the MCS$t$FA model introduced here is even more parsimonious than their most parsimonious model. One would therefore expect that the MCS$t$FA model will bring advantages in the analysis of very high dimensional data. Future work will study the point at which latent factor models fail and latent common factor models still succeed --- this is a bigger issue and will involve at least three cases: Gaussian mixtures, $t$-mixtures, and skew-$t$ mixtures. With high-dimensional applications in mind, future work will also focus on the efficient implementation of the mixture of skew-$t$ factor analyzers and MCS$t$FA models in parallel, building on the coarse-grain implementation of the PGMM family developed by \cite{mcnicholas10a}. Finally, it will also be interesting to develop a mixture of common skew-normal factor analyzers model and compare it to the MCFA model.

\section*{Acknowledgements}
This work was supported by an Early Researcher Award from the Ontario Ministry of Research and Innovation (McNicholas), and respective Discovery Grants from the Natural Sciences and Engineering Research Council of Canada (Browne, McNicholas). The authors are grateful to Professor Geoff McLachlan for providing access to the {\tt mixfa} package. 

\bibliographystyle{chicago}
\bibliography{MCSTFAref}







\end{document}